\documentstyle[aps,epsf]{revtex}

\begin{document}

\title{Anisotropic Stars in General Relativity}
\author{M. K. Mak\footnote{E-mail:mkmak@vtc.edu.hk}}
\address{Department of Physics, The Hong Kong University of Science and Technology,Clear Water Bay, Hong Kong, P. R. China.}
\author{T. Harko\footnote{E-mail: tcharko@hkusua.hku.hk}}
\address{Department of Physics, The University of Hong Kong,
Pokfulam Road, Hong Kong, P. R. China.}

\date{March 11, 2002}
\maketitle

\begin{abstract}

We present a class of exact solutions of Einstein's gravitational field
equations describing spherically symmetric and static anisotropic stellar
type configurations. The solutions are obtained by assuming a particular
form of the anisotropy factor. The energy density and both radial and
tangential pressures are finite and positive inside the anisotropic star.
Numerical results show that the basic physical parameters (mass and radius)
of the model can describe realistic astrophysical objects like neutron stars.



\end{abstract}

\section{Introduction}

Relativistic stellar models have been studied ever since the first solution
of Einstein's field equation for the interior of a compact object in
hydrostatic equilibrium was obtained by Schwarzschild in 1916. The search
for the exact solutions describing static isotropic and anisotropic stellar
type configurations has continuously attracted the interest of physicists.
The study of general relativistic compact objects is of fundamental
importance for astrophysics. After the discovery of pulsars and explanation
of their properties by assuming them to be rotating neutron stars, the
theoretical investigation of superdense stars has been done using both
numerical and analytical methods and the parameters of neutron stars have
been worked out by general relativistic treatment.

There are very few exact interior solutions (both isotropic and anisotropic)
of the gravitational field equations satisfying the required general
physical conditions inside the star. From $127$ published solutions analyzed
in Delgaty and Lake (1998) only $16$ satisfy all the conditions. But the
study of the interior of general relativistic stars via finding exact
solutions of the field equations is still an active field of research. Hence
recently an algorithm was proposed by Fodor (2000) to generate any number of
physically realistic pressure and density profiles for spherical perfect
fluid isotropic distributions without evaluating integrals. The
gravitational field equations for static stellar models with a linear
barotropic equation of state Nilsson and Uggla (2001b) and with a polytropic
equation of state Nilsson and Uggla (2001a) $p=k\rho ^{1+1/n}$ were recast
respectively into two complementary 3-dimensional regular system of ordinary
differential equation on compact state space, these systems were analyzed
numerically and qualitatively, using the theory of dynamical system. Schmidt
and Homann (2000) discussed numerical solutions of Einstein's field equation
describing static spherically symmetric conglomerations of a photon star
with an equation of state $\rho =3p$. Recently upper limits for the
mass-radius ratio are derived for compact general relativistic objects in
the presence of a cosmological constant (Mak, Dobson and Harko 2000) and in
the presence of a charge distribution (Mak, Dobson and Harko 2001). 

Since the pioneering work of Bowers and Liang (1974) there is an extensive
literature devoted to the study of anisotropic spherically symmetric static
general relativistic configurations. The theoretical investigations of
Ruderman (1972) about more realistic stellar models show that the nuclear
matter may be anisotropic at least in certain very high density ranges ($%
\rho >10^{15}g/cm^{3}$), where the nuclear interactions must be treated
relativistically. According to these views in such massive stellar objects
the radial pressure may not be equal to the tangential one. No celestial
body is composed of purely perfect fluid. Anisotropy in fluid pressure could
be introduced by the existence of a solid core or by the presence of type $%
3A $ superfluid (Kippenhahm and Weigert 1990), different kinds of phase
transitions (Sokolov 1980), pion condensation (Sawyer 1972) or by other
physical phenomena. On the scale of galaxies, (Binney and Tremaine 1987)
have considered anisotropies in spherical galaxies, from a purely Newtonian
point of view.

The mixture of two gases (e.g., monatomic hydrogen, or ionized hydrogen and
electrons) can formally be also described as an anisotropic fluid (Letelier
1980, Bayin 1982). Bowers and Liang (1974) have investigated the possible
importance of locally anisotropic equations of state for relativistic fluid
spheres by generalizing the equations of hydrostatic equilibrium to include
the effects of local anisotropy. Their study shows that anisotropy may have
non-negligible effects on such parameters as maximum equilibrium mass and
surface redshift. Heintzmann and Hillebrandt (1975) studied fully
relativistic, anisotropic neutron star models at high densities by means of
several simple assumptions and have shown that for arbitrary large
anisotropy there is no limiting mass for neutron stars, but the maximum mass
of a neutron star still lies beyond $3-4M_{\odot }$. Hillebrandt and
Steinmetz (1976) considered the problem of stability of fully relativistic
anisotropic neutron star models. They derived the differential equation for
radial pulsations and showed that there exists a static stability criterion
similar to the one obtained for isotropic models. Anisotropic fluid sphere
configurations have been analyzed, using various Ansatze, in Bayin (1982),
Cosenza, Herrera, Esculpi and Witten (1981) and Harko and Mak (2000). For
static spheres in which the tangential pressure differs from the radial one,
Bondi (1992) has studied the link between the surface value of the potential
and the highest occurring ratio of the pressure tensor to the local density.
Chan, Herrera and Santos (1993) studied in detail the role played by the
local pressure anisotropy in the onset of instabilities and they showed that
small anisotropies might in principle drastically change the stability of
the system. Herrera and Santos (1995) have extended the Jeans instability
criterion in Newtonian gravity to systems with anisotropic pressures. Recent
reviews on isotropic and anisotropic fluid spheres can be found in Delgaty
and Lake (1998) and Herrera and Santos (1997).

Very recently an analysis based on the Weyl tensor of the
Lemaitre-Schwarzschild problem of finding the equilibrium conditions of an
anisotropically sustained spherical body under its relativistic
gravitational field has been revisited by Fuzfa, Gerard and Lambert (2001).
Hernandez and Nunez (2001) presented a general method for obtaining static
anisotropic spherically symmetric solutions satisfying a nonlocal equation
of state from known density profiles, by assuming the condition of a
vanishing Weyl tensor. This condition can be integrated in the spherically
symmetric case. Then, the resulting expression is used by Herrera, Prisco,
Ospino and Fuenmayor (2001) to find new, conformally flat, interior
solutions to Einstein equations for locally anisotropic fluid. Dev and
Gleiser (2000) presented several exact solutions for anisotropic stars of
constant density. Some classes of exact solutions of Einstein field
equations describing spherically and static anisotropic stellar type
configurations were presented by Harko and Mak (2002) and Mak, Dobson and Harko (2001).
The surface redshift $z_s$ of anisotropic realistic stars has been investigated by Ivanov (2002).
$z_s$ cannot exceed the values $3.842$ or $5.211$, when the tangential
pressure satisfies the strong or dominant energy conditions, respectively.

In the present paper we consider a class of exact solutions of the
gravitational field equations for an anisotropic fluid sphere, corresponding
to a specific choice of the anisotropy parameter. The metric functions can
be represented in a closed form in terms of elementary functions. All the
physical parameters like the energy density, pressure and metric tensor
components are regular inside the anisotropic star, with the speed of sound
less than the speed of light. Therefore this solution can give a
satisfactory description of realistic astrophysical compact objects like
neutron stars. Some explicit numerical models of relativistic anisotropic
stars, with a possible astrophysical relevance, are also presented.

This paper is organized as follows. In Section II we present an exact class
of solutions for an anisotropic fluid sphere. In Section III we present
neutron star models with possible astrophysical applications. The results
are summarized and discussed in Section IV.

\section{Non-Singular Models for Anisotropic Stars}

In standard coordinates $x^{i}=\left( t,r,\chi ,\phi \right) $, the general
line element for a static spherically symmetric space-time takes the form 
\begin{equation}  \label{(2.1)}
ds^{2}=A^{2}(r)dt^{2}-V^{-1}(r)dr^{2}-r^{2}\left( d\chi ^{2}+\sin ^{2}\chi
d\phi ^{2}\right) .  \eqnum{2.1}
\end{equation}

Einstein's gravitational field equations are (where natural units $8\pi
G=c=1 $ have been used throughout): 
\begin{equation}
R_{i}^{k}-\frac{1}{2}R\delta _{i}^{k}=T_{i}^{k}.  \eqnum{2.2}  \label{(2.2)}
\end{equation}

For an anisotropic spherically symmetric matter distribution the components
of the energy-momentum tensor are of the form 
\begin{equation}
T_{i}^{k}=\left( \rho +p_{\perp }\right) u_{i}u^{k}-p_{\perp }\delta
_{i}^{k}+\left( p_{r}-p_{\perp }\right) \chi _{i}\chi ^{k},  \eqnum{2.3}
\label{(2.3)}
\end{equation}
where $u^{i}$ is the four-velocity $Au^{i}=\delta _{0}^{i}$, $\chi ^{i}$ is
the unit spacelike vector in the radial direction $\chi ^{i}=\sqrt{V}\delta
_{1}^{i}$, $\rho $ is the energy density, $p_{r}$ is the pressure in the
direction of $\chi ^{i}$ (normal pressure) and $p_{\perp }$ is the pressure
orthogonal to $\chi _{i}$ (transversal pressure). We assume $p_{r}\neq
p_{\perp }$. The case $p_{r}=p_{\perp }$ corresponds to the isotropic fluid
sphere. $\Delta =p_{\perp }-p_{r}$ is a measure of the anisotropy and is
called the anisotropy factor Herrera and Ponce de Leon (1985).

A term $\frac{2\left( p_{\perp }-p_{r}\right) }{r}$ appears in the
conservation equations $T_{k;i}^{i}=0$, (where a semicolon $;$ denotes the
covariant derivative with respect to the metric), representing a force that
is due to the anisotropic nature of the fluid. This force is directed
outward when $p_{\perp }>p_{r}$ and inward when $p_{\perp }<p_{r}$. The
existence of a repulsive force (in the case $p_{\perp }>p_{r}$) allows the
construction of more compact objects when using anisotropic fluid than when
using isotropic fluid (Gokhroo and Mehra 1994).

For the metric (2.1) the gravitational field equations (2.2) become 
\begin{equation}
\rho =\frac{1-V}{r^{2}}-\frac{V^{\prime }}{r},p_{r}=\frac{2A^{\prime }V}{Ar}+%
\frac{V-1}{r^{2}},  \eqnum{2.4}  \label{(2.4)}
\end{equation}
\begin{equation}
V^{\prime }\left( \frac{A^{\prime }}{A}+\frac{1}{r}\right) +2V\left( \frac{%
A^{\prime \prime }}{A}-\frac{A^{\prime }}{rA}-\frac{1}{r^{2}}\right)
=2\left( \Delta -\frac{1}{r^{2}}\right) ,  \eqnum{2.5}  \label{(2.5)}
\end{equation}
where $\prime =\frac{d}{dr}$.

To simplify calculations it is convenient to introduce the following
substitutions: 
\begin{equation}
V=1-2x\eta ,x=r^{2},\eta \left( r\right) =\frac{m(r)}{r^{3}},m(r)=\frac{1}{2}%
\int_{0}^{r}\xi ^{2}\rho \left( \xi \right) d\xi .  \eqnum{2.6}
\label{(2.6)}
\end{equation}

$m(r)$ represents the total mass content of the distribution within the
fluid sphere of radius $r$. Hence, we can express (2.5) in the equivalent
form 
\begin{equation}
\left( 1-2x\eta \right) \frac{d^{2}A}{dx^{2}}-\left( x\frac{d\eta }{dx}+\eta
\right) \frac{dA}{dx}-\left( \frac{1}{2}\frac{d\eta }{dx}+\frac{\Delta }{4x}%
\right) A=0.  \eqnum{2.7}  \label{(2.7)}
\end{equation}

For any physically acceptable stellar models, we require the condition that
the energy density is positive and finite at all points inside the fluid
spheres. In order to have a monotonic decreasing energy density $\rho =\frac{%
2}{r^{2}}\frac{d}{dr}\left( \eta r^{3}\right) $ inside the star we chose the
function $\eta $ in the form 
\begin{equation}
\eta =\frac{a_{0}}{2\left( 1+\psi \right) },  \eqnum{2.8}  \label{(2.8)}
\end{equation}
where $\psi =c_{0}x$, and $a_{0}$, $c_{0}$ are non-negative constant. (2.8) represents an ansatz
for the mass function, which has been previously used in the study of the isotropic
fluid spheres by Matese and Whitman (1980) and Finch and Skea (1989).

We also introduce a new variable $\lambda $ by means of the transformation 
\begin{equation}
\lambda =\frac{\left( a_{0}-c_{0}\right) \left( 1+\psi \right) }{a_{0}}. 
\eqnum{2.9}  \label{(2.9)}
\end{equation}

We also chose the anisotropy parameter as 
\begin{equation}
\Delta =\frac{c_{0}\Delta _{0}\psi }{\left( \psi +1\right) ^{2}}, 
\eqnum{2.10}  \label{(2.10)}
\end{equation}
where $a_{0}/c_{0}=\frac{\Delta _{0}-K}{1-K}$ or equivalently $\Delta
_{0}=a_{0}\left( 1-K\right) /c_{0}+K$ and $K$ a constant. In the following
we assume $\Delta _{0}\geq 0$, with $\Delta _{0}=0$ corresponding to the
isotropic limit. Hence $\Delta _{0}$ can be considered, in the present
model, as a measure of the anisotropy of the pressure distribution inside
the fluid sphere. At the center of the fluid sphere the anisotropy vanishes, 
$\Delta (0)=0$. For small $r$ the anisotropy parameter increases and, after
reaching a maximum in the interior of the star, it becomes a decreasing
function of the radial distance.

Therefore with this choice of $\Delta $, (2.7) becomes a hypergeometric
equation, 
\begin{equation}
\lambda \left( \lambda -1\right) \frac{d^{2}A}{d\lambda ^{2}}+\frac{1}{2}%
\frac{dA}{d\lambda }-\frac{K}{4}A=0.  \eqnum{2.11}  \label{(2.11)}
\end{equation}

By introducing the substitution $\lambda =1-X^{2}$, (2.11) reduces to a
standard second order differential equation given by 
\begin{equation}
\left( 1-X^{2}\right) \frac{d^{2}A}{dX^{2}}+X\frac{dA}{dX}+KA=0. 
\eqnum{2.12}  \label{(2.12)}
\end{equation}

By means of the pair of transformations (Gupta and Jasim 2000) 
\begin{equation}
\frac{dA}{dX}=G,X=\sin \theta ,  \eqnum{2.13}  \label{(2.13)}
\end{equation}
(2.12) reads 
\begin{equation}
\frac{d^{2}G}{d\theta ^{2}}+\omega ^{2}G=0,  \eqnum{2.14}  \label{(2.14)}
\end{equation}
where $\omega ^{2}=1+K$.

We shall not present here a simple linear solution for $G$ corresponding to $%
\omega ^{2}=0$. For nonzero $\omega ^{2}$ and with the use of the general solution
of (2.14), (2.13) becomes
\begin{equation}
\frac{dA}{dX}=G=p_{1}\cos \left( \omega \sin ^{-1}X\right) +q_{1}\sin \left( \omega \sin
^{-1}X\right),  \eqnum{2.15}  \label{(2.15)}
\end{equation}
where $p_1$ and $q_1$ are arbitrary constants of integration.

In view of (2.12) and (2.15) and after some re-adjustments of the arbitrary
constants, we obtain the general solution of (2.15) in the form 
\begin{eqnarray}
A &=&\alpha _{1}\left[ \omega \sqrt{1-X^{2}}\cos \left( \omega \sin ^{-1}X\right) +X\sin
\left( \omega \sin ^{-1}X\right) \right] +  \eqnum{2.16}  \label{(2.16)} \\
&&+\beta _{1}\left[ \omega \sqrt{1-X^{2}}\sin \left( \omega \sin ^{-1}X\right) -X\cos
\left( \omega \sin ^{-1}X\right) \right],  \nonumber
\end{eqnarray}
where $\alpha _{1}$ and $\beta _{1}$ are arbitrary constants of integration.

For simplicity, the metric functions and the physical parameters $\rho
,p_{\perp },p_{r}$ and $\Delta $ can be expressed in terms of $\theta $ by
setting $\sin \theta =X$, so $\lambda =\left( \frac{\Delta _{0}-1}{\Delta
_{0}-K}\right) \left( 1+\psi \right) =1-X^{2}=\cos ^{2}\theta $.

Therefore the general solution of the gravitational field equations for a
static anisotropic fluid sphere can be presented in the following form
expressed in elementary functions: 
\begin{equation}
A=\sqrt{C}\left\{ \left( \omega -1\right) \sin \left[ \left( \omega +1\right) \theta +B%
\right] +\left( \omega +1\right) \sin \left[ \left( \omega -1\right) \theta +B\right]
\right\},  \eqnum{2.17}  \label{(2.17)}
\end{equation}
\begin{equation}
V=1-\left( \frac{\Delta _{0}-K}{1-K}\right) \left( \frac{\psi }{1+\psi }%
\right) =\left( \frac{\Delta _{0}-1}{1-K}\right) \tan ^{2}\theta , 
\eqnum{2.18}  \label{(2.18)}
\end{equation}
\begin{equation}
\Delta =\frac{c_{0}\Delta _{0}\psi }{\left( 1+\psi \right) ^{2}}=\frac{%
c_{0}\Delta _{0}\left[ \left( \frac{\Delta _{0}-K}{\Delta _{0}-1}\right)
\cos ^{2}\theta -1\right] }{\left[ \left( \frac{\Delta _{0}-K}{\Delta _{0}-1}%
\right) \cos ^{2}\theta \right] ^{2}},  \eqnum{2.19}  \label{(2.19)}
\end{equation}
\begin{equation}
\rho =c_{0}\left( \frac{\Delta _{0}-K}{1-K}\right) \left[ \frac{3+\psi }{%
\left( 1+\psi \right) ^{2}}\right] =\frac{c_{0}\left( \Delta _{0}-1\right)
^{2}}{\left( 1-K\right) \left( \Delta _{0}-K\right) }\left[ \left( \frac{%
\Delta _{0}-K}{\Delta _{0}-1}\right) \sec ^{2}\theta +2\sec ^{4}\theta %
\right],  \eqnum{2.20}  \label{(2.20)}
\end{equation}
\begin{equation}
p_{r}=c_{0}\left( \frac{1-\Delta _{0}}{1-K}\right) \sec ^{2}\theta \left\{
2K\left( \frac{1-\Delta _{0}}{K-\Delta _{0}}\right) \left[ \frac{\tan \theta 
}{\omega \tan \left( \omega \theta +B\right) -\tan \theta }\right] +1\right\}, 
\eqnum{2.21}  \label{(2.21)}
\end{equation}
\begin{equation}
p_{\perp }=\frac{c_{0}\Delta _{0}\psi }{\left( 1+\psi \right) ^{2}}+p_{r}, 
\eqnum{2.22}  \label{(2.22)}
\end{equation}
where $\alpha _{1}=2\sqrt{C}\sin B$ and $\beta _{1}=2\sqrt{C}\cos B$ and $B$
and $C$ are arbitrary constants.

The physical quantities $\frac{dp_{r}}{d\rho }$ and $\frac{dp_{\perp }}{%
d\rho }$ describing the behaviour of the speed of sound inside the static
anisotropic fluid sphere can be computed from the resulting line element.

In order to be physically meaningful, the interior solution for static fluid
spheres of Einstein's gravitational field equations must satisfy some
general physical requirements. The following conditions have been generally
recognized to be crucial for anisotropic fluid spheres Herrera and Santos
(1997):

a) the density $\rho $ and pressure $p_{r}$ should be positive inside the
star;

b) the gradients $\frac{d\rho }{dr}$, $\frac{dp_{r}}{dr}$ and $\frac{%
dp_{\perp }}{dr}$ should be negative;

c) inside the static configuration the speed of sound should be less than
the speed of light, i.e. $0\leq \frac{dp_{r}}{d\rho }\leq 1$ and $0\leq 
\frac{dp_{\perp }}{d\rho }\leq 1$;

d) a physically reasonable energy-momentum tensor has to obey the conditions 
$\rho \geq p_{r}+2p_{\perp }$ and $\rho +p_{r}+2p_{\perp }\geq 0$;

e) the interior metric should be joined continuously with the exterior
Schwarzschild metric, that is $A^{2}(a)=1-2u$, where $u=M/a$, $M$ is the
mass of the sphere as measured by its external gravitational field and $a$
is the boundary of the sphere;

f) the radial pressure $p_{r}$ must vanish but the tangential pressure $%
p_{\perp }$ may not vanish at the boundary $r=a$ of the sphere. However, the
radial pressure is equal to the tangential pressure at the center of the
fluid sphere.

By matching (2.18) on the boundary of the anisotropic sphere and by denoting 
$\psi _{a}=c_{0}a^{2}$, we obtain 
\begin{equation}
V_{anis}\left( a\right) =1-\left( \frac{\Delta _{0}-K}{1-K}\right) \left( 
\frac{\psi _{a}}{1+\psi _{a}}\right) =1-2u_{anis},  \eqnum{2.23}
\label{(2.23)}
\end{equation}
where generally $u$ denotes the mass-radius ratio of the isotropic or
anisotropic star and the suffix ''$a$'' represents the value of the physical
quantities at the vacuum boundary of the star $r=a$.

For the isotropic case, that is for $\Delta _{0}=0$, it is easy to show that 
$u_{iso}=\frac{K}{2\left( K-1\right) }\left( \frac{\psi _{a}}{1+\psi _{a}}%
\right) $. Therefore the mass-radius ratios for the anisotropic and
isotropic spheres are related in the present model by 
\begin{equation}
u_{anis}=\frac{K-\Delta _{0}}{K}u_{iso}.  \eqnum{2.24}  \label{(2.24)}
\end{equation}

Hence, the constants $C$ and $B$ appearing in the solution can be evaluated
from the boundary conditions. Thus, we obtain 
\begin{equation}
\psi _{a}=c_{0}a^{2}=\frac{2\left( 1-K\right) u_{iso}}{2\left( K-1\right)
u_{iso}-K}=\frac{2\left( 1-K\right) u_{anis}}{2\left( K-1\right)
u_{anis}-K+\Delta _{0}},  \eqnum{2.25}  \label{(2.25)}
\end{equation}
\begin{equation}
C=\left( 1-2u_{anis}\right) \left\{ \left( \omega -1\right) \sin \left[ \left(
\omega +1\right) \theta _{a}+B\right] +\left( \omega +1\right) \sin \left[ \left(
\omega -1\right) \theta _{a}+B\right] \right\} ^{-2}.  \eqnum{2.26}  \label{(2.26)}
\end{equation}

Using (2.21) and the boundary condition that the radial pressure $p_{r\text{ 
}}$ vanishes at the radius $r=a$ of the fluid sphere, we obtain 
\begin{equation}
\tan B=\frac{\left[ \frac{K+\Delta _{0}\left( 1-2K\right) }{\omega \left( \Delta
_{0}-K\right) }\right] \tan \theta _{a}-\tan \left( \omega \theta _{a}\right) }{1+%
\left[ \frac{K+\Delta _{0}\left( 1-2K\right) }{\omega \left( \Delta _{0}-K\right) }%
\right] \tan \left( \omega \theta _{a}\right) \tan \theta _{a}}.  \eqnum{2.27}
\label{(2.27)}
\end{equation}

By defining $\tan \left( \omega \theta _{a}\right) =\left[ \tan \left( \omega \theta
\right) \right] _{r=a}$, it is easy to show that 
\begin{equation}
\cos \theta _{a}=\sqrt{\frac{\left( \Delta _{0}-1\right) \left( 1+\psi
_{a}\right) }{\left( \Delta _{0}-K\right) }},  \eqnum{2.28}  \label{(2.28)}
\end{equation}
\begin{equation}
\tan \theta _{a}=\sqrt{\frac{\left( \Delta _{0}-K\right) }{\left( \Delta
_{0}-1\right) \left( 1+\psi _{a}\right) }-1},\tan \left( \omega \theta _{a}\right)
=\tan \left[ \omega \tan ^{-1}\left( \sqrt{\frac{\left( \Delta _{0}-K\right) }{%
\left( \Delta _{0}-1\right) \left( 1+\psi _{a}\right) }-1}\right) \right]. 
\eqnum{2.29}  \label{(2.29)}
\end{equation}

The mathematical consistency of (2.28) requires the condition $\left( \Delta _0-1\right)\left( \Delta _0-K\right)\geq0$
be satisfied.

In order to find general constraints for the anisotropy parameter $\Delta
_{0}$, $K$ and $u_{anis}$, we shall consider that the conditions $\rho
_{0}=\rho \left( 0\right) \geq 0$, $p_{0}=p(0)\geq 0$ and $R_{0}=R(0)=\rho
_{0}-3p_{0}\geq 0$ hold at the center of the fluid sphere. Subsequently the
parameters $\Delta _{0}$, $K$ and $u_{anis}$ should be restricted to obey
the following conditions: 
\begin{equation}
\rho _{0}=\frac{3\left( \Delta _{0}-K\right) }{1-K}\geq 0,  \eqnum{2.30}
\label{(2.30)}
\end{equation}
\begin{equation}
p_{0}=\frac{\left( \Delta _{0}-K\right) \sqrt{1+K}+4K\left( K-1\right) \cot B%
}{\left( K-1\right) \sqrt{1+K}}\geq 0,  \eqnum{2.31}  \label{(2.31)}
\end{equation}
\begin{equation}
R_{0}=\frac{6\left[ \left( K-\Delta _{0}\right) \sqrt{1+K}-2K\left(
K-1\right) \cot B\right] }{\left( K-1\right) \sqrt{1+K}}\geq 0,  \eqnum{2.32}
\label{(2.32)}
\end{equation}

Clearly, the condition of the non-negativity of the energy density is
satisfied only if $\left( \Delta _{0}-K\right)\left( 1-K\right)\geq0$. 

The general behavior of the functions $\rho _{0}\left( K,\Delta _{0}\right) $%
, $p_{0}\left( K,\Delta _{0},u_{iso}\right) $ and $R_{0}(K,\Delta
_{0},u_{iso})$ is represented in Fig.1.

\begin{figure}[h]
\epsfxsize=10cm
\centerline{\epsffile{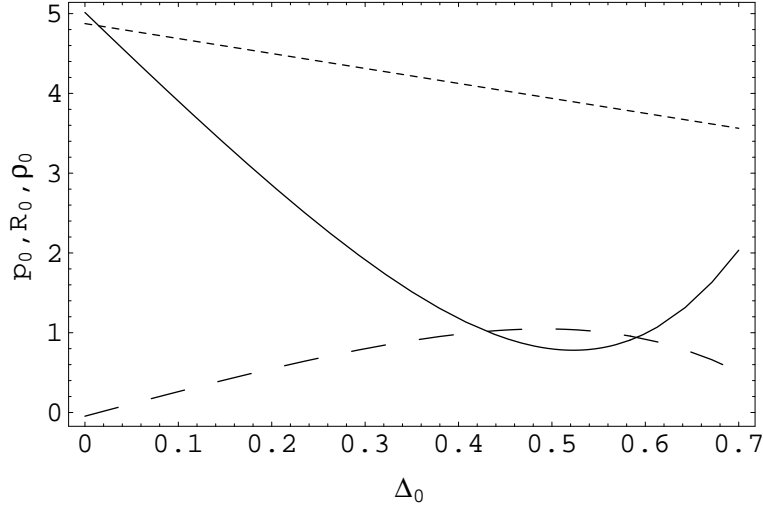}}
\caption{ Variations of the trace $R_{0}=\rho _{0}-3p_{0}$ of the
energy momentum tensor at the centre of the anisotropic star (solid curve),
of the central density $\rho _{0}$ (dotted curve) and of the central
pressure $p_{0}$ (dashed curve) as functions of the anisotropy parameter $%
\Delta _{0}$, for $K=2.6$ and for a given mass-radius ratio $u_{iso}=0.24$.}
\label{FIG1}
\end{figure}

Since $\cos \theta _{a}\leq 1$, from (2.28) it follows that the parameters $K
$, $\Delta _{0}$ and $\psi _{a}$ must satisfy the condition 
\begin{equation}
\frac{\left( \Delta _{0}-1\right) \left( 1+\psi _{a}\right) }{\left( \Delta
_{0}-K\right) }\leq 1.  \eqnum{2.33}
\end{equation}

(2.33) also represents the mathematical consistency condition for equations (2.29)
be well-defined.
The upper limit of the radius $a$ of the anisotropic star can be obtained from (2.33) in the form:
\begin{equation}
a\leq \sqrt{\frac{1-K}{c_{0}\left( \Delta _{0}-1\right) }}.  \eqnum{2.34}
\end{equation}

(2.34) requires for the parameters $K$ and $\Delta _{0}$ to satisfy the condition
$\left( 1-K\right)\left( \Delta _{0}-1\right)\geq0$. 

Since from (2.23) $\psi _{a}$ can be defined in terms of the total mass $M_{anis}$
as $\psi _{a}=\frac{2(1-K)M_{anis}/a}{\Delta _{0}-K-2\left( 1-K\right)
M_{anis}/a}$, (2.33) gives the minimum allowed mass of the anisotropic star
in the form
\begin{equation}
M_{anis}\geq \frac{\Delta _{0}-1}{2\left( 1-K\right) }c_{0}a^{3}. \eqnum{2.35}
\end{equation}

The conditions of the non-negativity of the energy density and maximum radius of the star give
the allowed ranges of the parameters $\Delta _0$ and $K$. There are two ranges
for these parameters making the physical requirements mathematically consistent, and
they are given by: (i) $K\leq 1$ and $\Delta _0\geq 1\geq K$ and (ii) $K\geq 1$
and $\Delta _0\leq 1\leq K$. For values of $\Delta _0$ and $K$ which do not belong to these
sets, the field equations do not have a physical solution. 

\section{Astrophysical Applications}

When the thermonuclear sources of energy in its interior are exhausted, a
spherical star begins to collapse under the influence of gravitational
interaction of its matter content. The mass energy continues to increase and
the star ends up as a compact relativistic cosmic object such as neutron
star, strange star or black hole. Important observational quantities for
such objects are the surface redshift, the central redshift and the mass and
radius of the star.

For a relativistic anisotropic star described by the solution presented in
the previous Section the surface redshift $z_{s}$ is given by 
\begin{equation}
z_{s}=\left( 1-2u_{anis}\right) ^{-1/2}-1.  \eqnum{3.1}  \label{(3.1)}
\end{equation}

The surface redshift is decreasing with increasing $\Delta _{0}$. Hence, at
least in principle, the study of redshift of light emitted at the surface of
compact objects can lead to the possibility of observational detection of
anisotropies in the internal pressure distribution of relativistic stars.

The central redshift $z_{c}$ is of the form 
\begin{equation}
z_{c}=C^{-1/4}\left\{ \left( \omega -1\right) \sin \left[ \left( \omega+1\right) \theta
_{0}+B\right] +\left( \omega-1\right) \sin \left[ \left( \omega+1\right) \theta _{0}+B%
\right] \right\} ^{-1/2}-1,  \eqnum{3.2}  \label{(3.2)}
\end{equation}
where $\theta _{0}=\cos ^{-1}\sqrt{\frac{\Delta _{0}-1}{\Delta _{0}-K}}.$
The variation of the central redshift of the neutron star against the
anisotropy parameter is represented in Fig. 2.

\begin{figure}[h]
\epsfxsize=10cm
\centerline{\epsffile{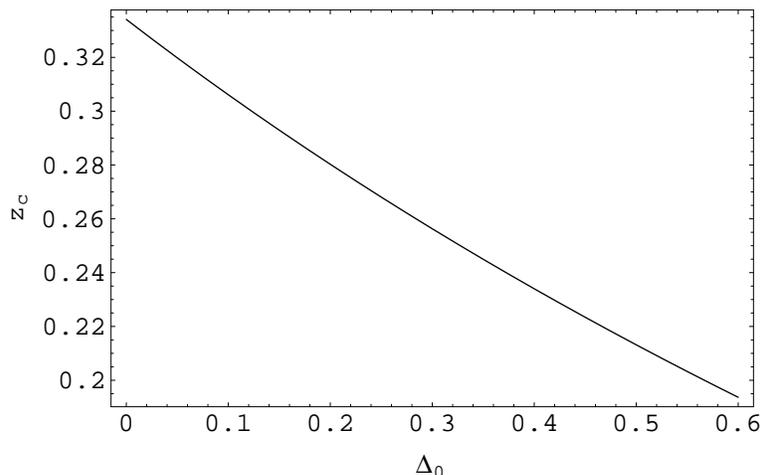}}
\caption{Variation of the central redshift $z_{c}$ as a function of the
anisotropy parameter $\Delta _{0}$ for the static anisotropic fluid sphere
with $K=2.2$ and a mass-radius ratio $u_{iso}=0.24$.}
\label{FIG2}
\end{figure}

Clearly, in our model the anisotropy introduced in the pressure gives rise
to a decrease in the central redshift. Hence, as functions of the
anisotropy, the central and surface redshifts have the same behavior.

The stellar model presented here can be used to describe the interior of the
realistic neutron star. Taking the surface density of the star as $\rho
_{s}=2\times 10^{14}g/cm^{3}$ and with the use of (2.25) we obtain 
\begin{equation}
\rho _{s}a_{N}^{2}=\frac{\left( \Delta _{0}-K\right) \psi _{a}\left( 3+\psi
_{a}\right) }{\left( 1-K\right) \left( 1+\psi _{a}\right) ^{2}},  \eqnum{3.3}
\label{(3.3)}
\end{equation}
or 
\begin{equation}
a_{N}=\frac{23.144}{K}\sqrt{\left( \Delta _{0}-K\right) u_{iso}\left[
4\left( K-1\right) u_{iso}-3K\right] }km,  \eqnum{3.4}  \label{(3.4)}
\end{equation}
where $a_{N}$ is the radius of the neutron star corresponding to a specific
surface density.

For the mass $M_{N}$ and anisotropy parameter $\Delta _{N}$ we find 
\begin{equation}
M_{N}=\frac{15.67}{K^{2}}\left( K-\Delta _{0}\right) u_{iso}\sqrt{\left(
\Delta _{0}-K\right) u_{iso}\left[ 4\left( K-1\right) u_{iso}-3K\right] }%
M_{\odot },  \eqnum{3.5}  \label{(3.5)}
\end{equation}
\begin{equation}
\Delta _{N}=3.59502\times 10^{35}\frac{\left( K-1\right) ^{2}\Delta
_{0}u_{iso}}{\left( \Delta _{0}-K\right) \left[ 4\left( K-1\right) u_{iso}-3K%
\right] }.  \eqnum{3.6}  \label{(3.6)}
\end{equation}

In the equations above we have expressed, for the sake of simplicity, all
the quantities in international units, instead of natural units, by means of
the transformations $\frac{M_{N}}{a_{N}}\rightarrow 8\pi \frac{GM_{N}}{%
c^{2}a_{N}}$, $\rho \rightarrow \rho c^{2}$ and $\Delta _{N}\rightarrow 
\frac{8\pi G}{c^{4}}\Delta _{N}$, where $G=6.6732\times 10^{-8}dyne{\ }cm^{2}%
{\ }g^{-2}$ and $c=2.997925\times 10^{10}cm{\ }s^{-1}$.

The variation of the anisotropy parameter $\Delta _{N}$ of the neutron star
as a function of the radius $a_{N}$ and mass $M_{N}$ is represented in Fig.
3.

\begin{figure}[h]
\epsfxsize=10cm
\centerline{\epsffile{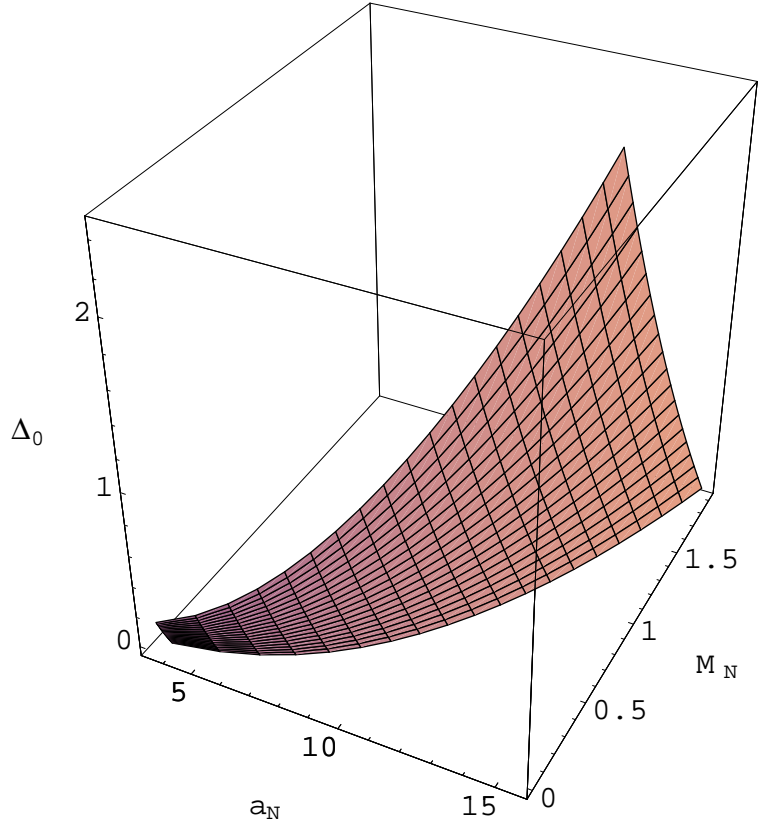}}
\caption{Variation of the anisotropy parameter of the neutron star $\Delta
_{N}$ (in units of $10^{4.7}$) as a function of the radius of the star $a_{N}
$ (in km) and of the mass $M_{N}$ of the star (in solar mass units), for the
mass-radius ratio $u_{iso}\in \left( 0.01,0.18\right) $ and $\Delta _{0}\in
\left( 0,0.6\right) $. We have used the value $K=2.2$.}
\label{FIG3}
\end{figure}

The quantities $\left( \frac{dp_{r}}{d\rho }\right) _{r=0}$, $\left( \frac{%
dp_{\perp }}{d\rho }\right) _{r=0}$ , $\left( \frac{dp_{r}}{d\rho }\right)
_{r=a_N}$ and $\left( \frac{dp_{\perp }}{d\rho }\right) _{r=a_N}$ are
represented against the anisotropy parameter $\Delta _{0}$ in Fig. 4.

\begin{figure}[h]
\epsfxsize=10cm
\centerline{\epsffile{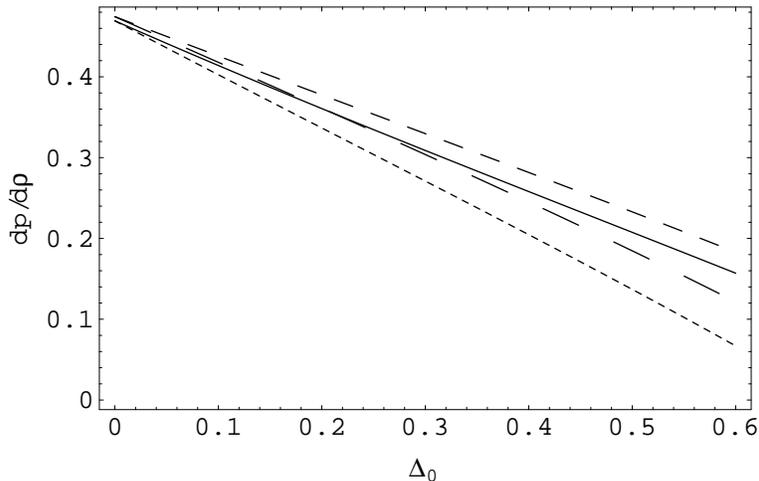}}
\caption{Variations of the radial and tangential speeds of sound $dp/d%
\rho $ at the vacuum boundary and at the center of the anisotropic
fluid sphere as a function of the anisotropy parameter $\Delta _{0}$ for $%
u_{iso}=0.2$ and $K=2.2$: $\left(dp_{r}/d\rho \right) _{r=0}$ (solid
curve), $\left(dp_{\perp }/d\rho \right) _{r=0}$ (dotted curve), $%
\left( dp_{r}/d\rho \right) _{r=a_N}$ (dashed curve) and $\left(
dp_{\perp }/d\rho \right) _{r=a_N}$ (long dashed curve).}
\label{FIG4}
\end{figure}

The plots indicate that the necessary and sufficient criterion for the
adiabatic speed of sound to be less than the speed of light is satisfied by
our solution. However, Caporaso and Brecher (1979) claimed that $dp/d\rho $
does not represent the signal speed. If therefore this speed exceeds the
speed of light, this does not necessary mean that the fluid is non-causal.
But this argument is quite controversial and not all authors accept it Glass
(1983).

\section{Discussions and Final Remarks}

In the present paper we have presented the general solution of the
gravitational field equations for an anisotropic static matter distribution.
To derive the solution we have used two basic assumptions: the functional
form of the mass function $\eta (r)$ and a specific mathematical
representation of the anisotropy parameter $\Delta $. In our model the
anisotropy vanishes at the center of the star, reaching its maximum near the
vacuum boundary.This behavior is similar, from a phenomenological point of
view, with the behavior of the anisotropy parameter of boson stars (Dev 
\& Gleiser 2001). Boson stars are gravitationally
bound  macroscopic quantum states made of scalar bosons (Jetzer 1992). They
differ from the fermionic stars in that they are only prevented from
collapsing gravitationally by the Heisenberg uncertainty principle. Boson
stars, described by non-interacting, massive scalar fields $\phi $ with
non-zero spatial gradients are anisotropic objects, with $\Delta =p_{\perp
}-p_{r}=-\left( \frac{d\phi }{dr}\right) ^{2}$. Hence our $\Delta $ has the
same qualitative properties as the anisotropy parameter of boson stars.

An other possible source of the anisotropy as described by (2.10) could be
an anisotropic velocity distribution of the particles inside the star. Most
of the models of stellar structure assume that the equation of state is due
to a single species of ideal noninteracting fermions, with an isotropic
distribution of momenta. If, due to the presence of a magnetic field,
convection, turbulence etc. the local velocity distribution is anisotropic,
then the pressure distribution is also anisotropic. To evaluate this effect
let us consider a spherical configuration of collisionless particles, which
exhibit no net velocity flow. The radial, azimuthal and polar velocities are 
$v_{r}$, $v_{\phi }$ and $v_{\theta }$, respectively. Then one can show that
for a system in Newtonian equilibrium $\Delta =p_{\perp }-p_{r}=\rho \left[
\left\langle v_{\theta }^{2}\right\rangle -\left\langle
v_{r}^{2}\right\rangle \right] $, where the brackets denote averages over
the particle distribution (Herrera \& Santos 1997).
Moreover, $\left\langle v_{\theta }^{2}\right\rangle -\left\langle
v_{r}^{2}\right\rangle =-\Phi \left\langle v_{r}^{2}\right\rangle $ (Herrera 
\& Santos 1997), where $\Phi $ is a parameter
measuring the anisotropy of velocity distributions, which for simplicity we
assume to be a constant. Therefore 
\begin{equation}
\Delta =\rho \left[ \left\langle v_{\theta }^{2}\right\rangle -\left\langle
v_{r}^{2}\right\rangle \right] =-c_{0}\left( \frac{\Delta _{0}-K}{1-K}%
\right) \left[ \frac{3+\psi }{\left( 1+\psi \right) ^{2}}\right] \Phi
\left\langle v_{r}^{2}\right\rangle.\eqnum{4.1} 
\end{equation}

But with the use of the virial theorem (the kinetic energy of particles is
equal to minus half of their potential energy) we obtain $\left\langle
v_{r}^{2}\right\rangle \sim -m/r$, where $m$ is the mass of the star.
Approximating the density by a constant (at least locally this is generally
a good approximation for general relativistic compact objects) it follows
that  $\left\langle v_{r}^{2}\right\rangle \sim -r^{2}\sim -\psi $.
Substituting this expression of the average of the square of the radial
velocity in (3.1), by chosing an appropriate value for the constant $\Phi $,
and by neglecting in the numerator the second order term proportional to $%
\psi ^{2}$ ($\psi \sim r^{2}/a^{2}<<1$ for  $r<a$), we obtain the exact form
of the anisotropy ansatz (2.10).            

The scalar invariants of the Riemann tensor are important since they allow a
manifestly coordinate invariant characterization of certain geometrical
properties of geometries, like for example curvature singularities, Petrov
type of the Weyl tensor etc. Two scalars which have been extensively used in
the physical literature are the Kretschmann scalars $RiemSq\equiv
R_{ijkl}R^{ijkl}=
\left( 2\Delta +\rho +p_{r}-\frac{4m}{r^{3}}\right) ^{2}+2\left( p_{r}+%
\frac{2m}{r^{3}}\right) ^{2}+2\left( \rho -\frac{2m}{r^{3}}\right)
^{2}+4\left( \frac{2m}{r^{3}}\right) ^{2}$
and $RicciSq\equiv R_{ij}R^{ij}=\rho ^{2}+3p_{r}^{2}+2\Delta \left( \Delta +2p_{r}\right) $,
where $R_{ijkl}$ is the Riemann curvature tensor.

For space-times which are the product of two 2-dimensional spaces, one
Lorentzian and one Riemannian, subject to a separability condition on the
function which couples the 2-spaces, Santosuosso et al. (1998) have
suggested that the set
\begin{equation}
C=\left\{ R,r_{1},r_{2},w_{2}\right\} , \eqnum{4.2}
\end{equation}
form an independent set of scalar polynomial invariants, satisfying the
number of degrees of freedom in the curvature. In (4.2) $%
R=g^{il}g^{jk}R_{ijkl}$ is the Ricci scalar and the quantities $r_{1}$, $%
r_{2}$ and $w_{2}$ are defined according to (Zakhary \& Carminati 2001; Carminati, Zakhary \& McLenaghan 2002)
\begin{equation}
r_{1}=\phi _{AB\dot{A}\dot{B}}\phi ^{AB\dot{A}\dot{B}}=\frac{1}{4}S_{a}^{b}S_{b}^{a},  \eqnum{4.3}
\end{equation}
\begin{equation}
r_{2}=\phi _{AB\dot{A}\dot{B}}\phi _{C\dot{C}}^{B\dot{B}}\phi ^{CA\dot{C}%
\dot{A}}=-\frac{1}{8}S_{a}^{b}S_{b}^{c}S_{c}^{a},   \eqnum{4.4}
\end{equation}
\begin{equation}
w_{2}=\Psi _{ABCD}\Psi _{EF}^{CD}\Psi ^{EFAB}=-\frac{1}{8}\bar{C}_{abcd}\bar{C}_{ef}^{cd}\bar{C}^{efab}, \eqnum{4.5}
\end{equation}
where $S_{a}^{b}=R_{a}^{b}-\frac{1}{4}R\delta _{a}^{b}$ is the trace-free
Ricci tensor, $\phi _{AB\dot{A}\dot{B}}$ denotes the spinor equivalent of $%
S_{ab}$, $\Psi _{ABCD}$ denotes the spinor equivalent of the Weyl tensor $%
C_{abcd}$ and $\bar{C}_{abcd}$ denotes the complex conjugate of the
self-dual Weyl tensor, $C_{abcd}^{+}=\frac{1}{2}\left( C_{abcd}-i\ast
C_{abcd}\right) $. 

In terms of the ''electric'' $E_{ac}=C_{abcd}u^{b}u^{d}$ and ''magnetic'' $%
H_{ac}=C_{abcd}^{\ast }u^{b}u^{d}$ parts of the Weyl tensor, where $u^{a}$
is a timelike unit vector and $C_{abcd}^{\ast }=\frac{1}{2}\eta
_{abef}C_{cd}^{ef}$ is the dual tensor, the invariant $w_{2}$ is given by
(Santosuosso et al. 1998)
\begin{equation}
w_{2}=\frac{1}{32}\left(
3E_{b}^{a}H_{c}^{b}H_{a}^{c}-E_{b}^{a}E_{c}^{b}E_{a}^{c}\right) +\frac{i}{32}%
\left( H_{b}^{a}H_{c}^{b}H_{a}^{c}-3E_{b}^{a}E_{c}^{b}H_{a}^{c}\right). \eqnum{4.6}
\end{equation}

For the case of a perfect fluid with anisotropic pressure distribution the
set of invariants (4.2) is given by:
\begin{equation}
R=\rho -3p_{r}-2\Delta , \eqnum{4.7}
\end{equation}
\begin{equation}
r_{1}=\frac{1}{64}\left[ 9\left( \rho +p_{r}+\frac{2\Delta }{3}\right)
^{2}+\left( \rho +p_{r}-2\Delta \right) ^{2}+2\left( \rho +p_{r}+2\Delta
\right) ^{2}\right] ,  \eqnum{4.8}
\end{equation}
\begin{equation}
r_{2}=-\frac{1}{512}\left[ 27\left( \rho +p_{r}+\frac{2\Delta }{3}\right)
^{3}+\left( \rho +p_{r}-2\Delta \right) ^{3}+2\left( \rho +p_{r}+2\Delta
\right) ^{3}\right],  \eqnum{4.9}
\end{equation}
\begin{eqnarray}
8\left(6^{\frac{1}{3}}\right)\left(w_{2}\right)^{2/3} &=&\left( \rho +p_{r}+2\Delta -\frac{4m}{r^{3}%
}\right) ^{2}+2\left( p_{r}+\frac{2m}{r^{3}}\right) ^{2}+2\left( \rho -\frac{%
2m}{r^{3}}\right) ^{2}+4\left( \frac{2m}{r^{3}}\right) ^{2}-\frac{1}{6}%
\left( \rho -3p_{r}-2\Delta \right) ^{2}- \nonumber \\
&&\frac{9}{8}\left( \rho +p_{r}+\frac{2\Delta }{3}\right) ^{2}-\frac{1}{8}%
\left( \rho +p_{r}-2\Delta \right) ^{2}-\frac{1}{4}\left( \rho
+p_{r}+2\Delta \right) ^{2}. \eqnum{4.10}
\end{eqnarray}

When $\Delta =0$ and $p_r=p_{\perp }=p$, we recover the form of the Ricci invariants
$r_n=c_n(\rho +p)^{n+1}$, $c_n=$constant, for class $B$ warped product space-times with
isotropic perfect fluid matter sources (Santosuosso et al. 1998).

The variations of $R$, $r_{1}$, $r_{2}$ and of $w_2$ at the center of the fluid sphere are represented, as a function
of the anisotropy parameter $\Delta _0$, in Fig. 5.

\begin{figure}[h]
\epsfxsize=10cm
\centerline{\epsffile{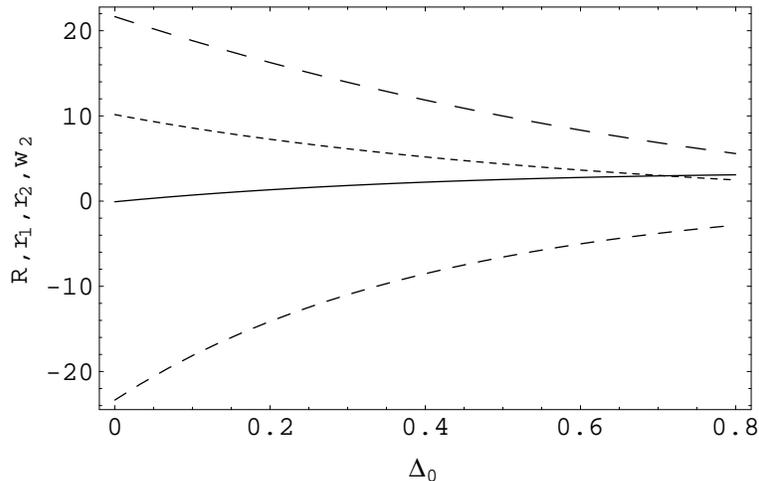}}
\caption{Variations, at the center $r=0$ of the static anisotropic fluid sphere, of the curvature scalar $R$ (solid curve), of
the Ricci invariants $r_{1}$ (dotted curve), $r_{2}$ (dashed curve) and of $w_2$ (long dashed curve)
against the anisotropy parameter $\Delta_{0}$, for $u_{iso}=0.29$ and $K=2.2$.}
\label{FIG5}
\end{figure}

The scalars $r_{1}$, $r_{2}$ and $w_2$ are finite at the center of the fluid sphere.
$r_{1}$ and $w_2$ monotonically decrease, while $r_{2}$ monotonically increases as the
anisotropy parameter $\Delta _{0}$ increases.

For a particular choice of the equation of state at the center of the star, $%
3p_{r0}=3p_{\perp 0}=\rho _{0}$ and with a vanishing anisotropy parameter, $%
\Delta _{0}=0$, we obtain the result $u_{iso}=0.29$ for a static isotropic
fluid sphere.

It is generally held that the trace $T$ of the energy-momentum tensor must
be non-negative. It is also the case that this trace condition is everywhere
fulfilled if it is fulfilled at the center of the star Knutsen (1988).

The purpose of the present paper is to present some exact models of static
anisotropic fluid stars and to investigate their possible astrophysical
relevance. All the solutions we have obtained are non-singular inside the
anisotropic sphere, with finite values of the density and pressure at the
center of the star. Variations of the physical parameters mass, radius,
redshift and adiabatic speed of sound against the anisotropy parameter have
been presented graphically. Because it satisfies all the physical conditions
and requirements (a)-(f), the present model can be used to study the
interior of the anisotropic relativistic objects.

\section*{Acknowledgments}

We would like to thank to the two anonymous referees whose comments helped us to improve the manuscript.

\begin{center}

{\bf References}

\end{center}

Bayin, S. S. 1982 Anisotropic fluid spheres in general relativity {\it Phys.
Rev}. D {\bf 26}, 1262-1274.

Binney, J. \& Tremaine, J. S. 1987 {\it Galactic Dynamics}, Princeton
University Press, Princeton, New Jersey.

Bondi, H. 1992 Anisotropic spheres in general relativity {\it Month. Not. R.
Astron. Soc}. {\bf 259}, 365-368.

Bowers, R. L. \& Liang, E. P. T. 1974 Anisotropic spheres in general
relativity {\it Astrophys. J.} {\bf 188}, 657-665.

Caporaso, G. \& Brecher, K. 1979 Must ultrabaric matter be superluminal {\it %
Phys. Rev}. D {\bf 20}, 1823-1831.

Carminati, J., Zakhary E. \& McLenaghan, R. G. 2002 On the problem of algebraic completeness for the
invariants

of the Riemann tensor: II {\it J. Math. Phys.} {\bf 43} 492-507.

Chan, R., Herrera, L. \& Santos, N. O. 1993 Dynamical instability for
radiating anisotropic collapse, {\it Month. Not.

R. Astron. Soc.} {\bf 265}, 533-544.

Cosenza, M., Herrera, L., Esculpi, M. \& Witten, L. 1981 Some models of
anisotropic spheres in general relativity,

{\it J. Math. Phys.} {\bf 22}, 118-125.

Delgaty, M. S. R. \& Lake, K. 1998 Physical acceptability of isolated,
static, spherically symmetric, perfect fluid

solutions of Einstein's equations {\it Comput. Phys. Commun}.
{\bf 115}, 395-415.

Dev, K. \& Gleiser M. 2000 Anisotropic stars: exact solutions, astro-ph/0012265.

Finch, M. R. \& Skea, J. E. F. 1989 A realistic stellar model based on the ansatz
of Duorah and Ray {\it Class. Quantum

Grav.} {\bf 6}, 467-476.

Fodor, G. 2000 Generating spherically symmetric static perfect fluid
solutions gr-qc/0011040.

Fuzfa, A., Gerard J. M. \& Lambert, D. 2001 The Lemaitre-Schwarzschild
problem revisited, gr-qc/0109097.

Glass, E. N. 1983 Ultrabaric neutron stars are superluminal {\it Phys. Rev}.
D {\bf 28}, 2693.

Gokhroo, M. K. \& Mehra, A. L. 1994 Anisotropic spheres with variable energy
density in general relativity, {\it Gen.

Rel. Grav.} {\bf 26}, 75-84.

Gupta, Y. K. \& Jasim, M. K. 2000 On most general exact solution for
Vaidya-Tikekar isentropic superdense star

{\it , Astrophys. Space Sci}. {\bf %
272}, 403-415.

Harko, T. \& Mak, M. K. 2000 Anisotropic charged fluid spheres in D
space-time dimensions {\it J. Math. Phys.} {\bf 41},

4752-4764.

Harko T. \& Mak M. K. 2002 Anisotropic relativistic stellar models {\it %
Annalen der Physik}, {\bf 11}, 3-13.

Heintzmann, H. \& Hillebrandt, W. 1975 Neutron stars with an anisotropic
equation of state: mass, redshift and

stability {\it Astron. Astrophys.} {\bf 38}, 51-55.

Herrera, L. \& Ponce de Leon J. 1985 Isotropic and anisotropic charged
spheres admitting a one-parameter group

of conformal motions {\it J. Math. Phys}. {\bf 26}, 2302-2307.

Herrera, L., Prisco A Di, Ospino J. \& Fuenmayor E. 2001 Conformally flat
anisotropic spheres in general relativity

{\it J. Math. Phys}. {\bf 42}, 2129-2143.

Herrera, L. \& Santos, N. O. 1995 Jeans mass for anisotropic matter {\it %
Astrophys. J}. {\bf 438}, 308-313.

Herrera, \ L. \& \ Santos, N. O. 1997 Local anisotropy in self-gravitating
systems {\it Phys. Reports} {\bf 286}, 53-130.

Hernandez H. \& Nunez L. A. 2001 Nonlocal equation of state in anisotropic
static fluid spheres in general relativity,

gr-qc/0107025.

Hillebrandt, W. \& Steinmetz, K. O. 1976 Anisotropic neutron star
models:Stability against radial and nonradial

pulsations {\it Astron. Astrophys}. {\bf 53},
283-287.

Ivanov, B. V. 2002 Maximum bounds on the surface redshift of anisotropic stars {\it gr-qc/0201090}, to appear in {\it Phys.

Rev. D}.

Jetzer, P. 1992 Boson stars {\it Phys. Reports} {\bf 220}, 163-227.

Kippenhahm, R. \& Weigert, A. 1990 {\it Stellar Structure and Evolution},
Springer, Berlin.

Knutsen, H. 1988 On the stability and physical properties of an exact
relativistic model for a superdense star {\it Month.

Not. R. Astr. Soc}. {\bf 232}, 163-174.

Letelier, P. 1980 Anisotropic fluids with two-perfect-fluid components {\it %
Phys. Rev}. D {\bf 22}, 807-813.

Mak M. K., Dobson P. N. jr. \& Harko T. 2000 Maximum mass-ratio ratios for
compact general relativistic objects

in Schwarzschild-de Sitter {\it Mod. Phys. Lett.
A} {\bf 15}, 2153-2158.

Mak M. K., Dobson P. N. jr. \& Harko T. 2001 Maximum mass-ratio ratios for
charged compact general relativistic

objects {\it Europhys. Lett}.{\it \ } {\bf 55}, 310-316.

Mak M. K., Dobson P. N. jr. \& Harko T. 2002 Exact model for anisotropic
relativistic stars. {\it Int. J. Mod. Phys.

D} {\bf 11}, 207-221. 

Matese J. J. \& Whitman P. G. 1980 New method for extracting equilibrium configurations
in general relativity

{\it Phys. Rev. D} {\bf 22}, 1270-1275.

Nilsson U. S. \& Uggla C. 2001a General relativistic stars: polytropic
equations of state {\it Annals Phys}. {\bf 286}, 292-319.

Nilsson U. S. \& Uggla C. 2001b General relativistic stars: linear equations
of state {\it Annals Phys}. {\bf 286}, 278-291.

Ruderman, R. 1972 Pulsars: Structure and dynamics {\it Ann. Rev. Astron.
Astrophys}. {\bf 10}, 427-476.

Santusuosso K., Pollney D., Pelavas N., Musgrave P. \& Lake, K. 1998 Invariants of the Riemann tensor for
class B

warped product space-times {\it Comp. Phys. Comm.} {\bf 115}, 381-394.

Sawyer, R. F. 1972 Condensed $\pi ^{-}$ phase in neutron-star matter {\it %
Phys. Rev. Lett}. {\bf 29}, 382-385.

Schmidt H. J. \& Homann F. 2000 Photon stars {\it Gen. Rel. Grav}. {\bf 32},
919-931.

Sokolov, A. I. 1980 Phase transitions in a superfluid neutron fluid {\it JETP%
} {\bf 79}, 1137-1140.

Zakhary E. \& Carminati, J. 2001 On the problem of algebraic completeness for the
invariants of the Riemann

tensor: I {\it J. Math. Phys.} {\bf 42} 1474-1485.

\end{document}